\documentclass[conference]{IEEEtran}
\usepackage{cite}
\usepackage{caption}
\usepackage{subfiles}
\usepackage{amsmath,amssymb,amsfonts}
\usepackage{algpseudocode}
\usepackage{algorithm}
\usepackage{graphicx}
\usepackage{textcomp}
\usepackage{epsfig}
\usepackage{makecell}

\usepackage{multirow}
\usepackage{steinmetz}
\usepackage{float}
\usepackage{subfig}

\newcommand{\R}[1]{\mathbb{R}^{#1}}
\newcommand{\T}[1]{{\boldsymbol{\mathcal{#1}}}}

\usepackage{booktabs}
\def\BibTeX{{\rm B\kern-.05em{\sc i\kern-.025em b}\kern-.08em
		T\kern-.1667em\lower.7ex\hbox{E}\kern-.125emX}}
\begin{document}

	\title{A divide-and-conquer approach for \\sparse recovery of high dimensional signals\\
	}
	
 \author{Aron Bevelander, Kim Batselier, Nitin Jonathan Myers \\
		Delft Center for Systems and Control, Delft University of Technology, The Netherlands\\
		 Email: aronadam97@gmail.com, \{K.Batselier, N.J.Myers\}@tudelft.nl}
	\maketitle

\begin{abstract} 
Compressed sensing (CS) techniques demand significant storage and computational resources, when recovering high-dimensional sparse signals. Block CS (BCS), a special class of CS, addresses both the storage and complexity issues by partitioning the sparse recovery problem into several sub-problems. In this paper, we derive a Welch bound-based guarantee on the reconstruction error with BCS. Our guarantee reveals that the reconstruction quality with BCS monotonically reduces with an increasing number of partitions. To alleviate this performance loss, we propose a sparse recovery technique that exploits correlation across the partitions of the sparse signal. Our method outperforms BCS in the moderate SNR regime, for a modest increase in the storage and computational complexities. 
\end{abstract}
\section{Introduction }
\par Compressed sensing (CS) is a technique to recover sparse high-dimensional signals from their low-dimensional representation. The dimension of sparse signals in typical CS applications is rapidly increasing with advances in sensing technology. For instance, sparse signals in a million dimension space have been considered in the context of fluorescence microscopy \cite{microscopy}, imaging \cite{singlepixel}, and wireless channel estimation using terahertz arrays \cite{THz_channel}. Solving for such high-dimensional signals in real-time requires intense memory and computational resources that may often be impractical or unavailable.


\par Block compressed sensing (BCS), a class of CS, adopts a divide-and-conquer approach to reduce the storage and computational complexities needed for sparse recovery \cite{BCS_1}. In BCS, the sparse recovery problem is partitioned into sub-problems, where each sub-problem solves for a block within the sparse vector. The sub-problems have a lower complexity than the original problem and they can be solved in parallel. BCS was applied in \cite{BCS_permutation} to reduce the storage needed for sparse recovery. Standard BCS methods typically obtain an equal number of CS measurements for each block, assuming a uniform distribution of sparse non-zero components across the blocks. In \cite{BCS_adaptive_measuring}, however, varying numbers of measurements were acquired per block based on its expected sparsity level. In \cite{BCS_permutation}, a permutation technique was proposed to ``equally'' distribute the non-zero components over different blocks. An important question in BCS is if partitioning the problem impacts reconstruction performance. An empirical study conducted in \cite{BCS_recon} showed that partitioning results in a higher mean squared error in the reconstruction. To the best of our knowledge, however, the fundamental limits on reconstruction performance with BCS as a function of the number of partitions or blocks have not been studied.

\par To overcome the performance loss with standard BCS, the sub-problems of BCS can be solved in series. Assuming that the signal is correlated across the blocks, information from one reconstructed block can be used as a prior in reconstructing other correlated blocks. Prior work on dynamical CS has extensively looked at exploiting such correlation information in a different context than BCS. For instance, the techniques in \cite{DCS_AMP,DCS_core_asif} exploit temporal correlation in the signal by incorporating signal dynamics within the sparse recovery problem. Here, each temporal snapshot can be interpreted as a block within a high dimensional sparse signal. In \cite{IT_paper_LWOMP}, sparse recovery algorithms were developed to exploit known probabilistic information on the support of the sparse signal. These probabilities can potentially be derived from the correlated blocks within BCS. Sequentially solving the sub-problems in BCS by exploiting block correlations can enhance reconstruction performance, albeit at the cost of losing the ability to solve the sub-problems in parallel. 
\par In this paper, we study the fundamental limits of BCS and discuss how to exploit correlation across different blocks for sparse recovery. First, we derive a Welch bound-based guarantee on the reconstruction error with BCS. Then, we develop a data-driven approach to learn the correlation across blocks. The learned correlation is then used for sparse recovery in our serial BCS method. Finally, we consider a near-field channel estimation problem to show that our data-driven serial BCS technique outperforms standard BCS at a moderate SNR.

\par Notation: $a\in\R{}$ denotes a scalar, $\textbf{a}\in\R{n}$ a vector, $\textbf{A}\in\R{n_1\times n_2}$ a matrix and $\T{A}\in\R{n_1\times n_2\times\dots\times n_d}$ a $d$-th order tensor. The convolution of $\T{A}$ and $\T{B}\in\R{m_1\times m_2\times\dots\times m_d}$ is denoted by $\T{C}=\T{A}\circledast \T{B}\in\R{k_1\times k_2\times\dots\times k_d}$ where $k_i=\max(n_i,m_i)\;i\in\{1,2,\dots,d\}$, such that its entry $\T{C}(i_1,i_2,\ldots,i_d) = \sum_{j_1=1}^{n_1} \sum_{j_2=1}^{n_2} \ldots \sum_{j_d=1}^{n_d} \T{A}(j_1,j_2,\ldots,j_d)\T{B}(i_1-j_1,i_2-j_2,\ldots,i_d-j_d)$. $\langle \mathbf{a}, \mathbf{b} \rangle$ is the inner product and $\Vert \mathbf{a} \Vert_2$ is the $\ell_2$ norm of $\mathbf{a}$. $[N]$ denotes the set $\{1,2,\cdots N\}$. 
\section{Analysis of block compressed sensing}
\subsection{Motivation for BCS}
\par In classical CS, a sparse signal $\textbf{x}\in\R{n}$ is projected on an $m$-dimensional space using a CS matrix $\textbf{A}\in\R{m\times n}$ with $m\ll n$. The projections are usually perturbed by noise $\textbf{v}\in\R{m}$, where each entry has a variance of $\sigma^2$. The vector containing the noisy compressed representation of $\textbf{x}$ is defined as 
\begin{equation}
    \textbf{y}=\textbf{A}\textbf{x}+\textbf{v}.
    \label{eq:CS_meas_model}
\end{equation}
CS algorithms estimate a sparse vector that best explains the observed measurements. Computationally tractable algorithms based on $\ell_1$ norm minimization and greedy methods have been developed to solve for $\textbf{x}$ from \eqref{eq:CS_meas_model}. These algorithms can find a sparse solution when $\mathbf{x}$ is sufficiently sparse and when the CS matrix $\mathbf{A}$ satisfies the restricted isometry property or has a low mutual coherence \cite{ben2010coherence}. 
\par The computational complexity of CS algorithms in \cite{tropp2007signal,blumensath2009iterative, donoho2009message} is $\mathcal{O}(mn)$ for $o(1)$ sparse signals, due to matrix-vector multiplications with $\mathbf{A}$ and $\mathbf{A}^T$. For high dimensional signals, both $n$ and $m$ can be significant so that real-time sparse recovery becomes infeasible under practical constraints.

\subsection{Storage and computational complexity of BCS}
\par BCS uses a block-diagonal structure for the CS matrix to alleviate the heavy storage demanded by generic CS matrices. We define $\beta$ as the number of blocks or partitions in BCS, and assume that the blocks are of equal size. To model discontiguous blocks in BCS, we define a permutation matrix $\boldsymbol{\Pi}\in \mathbb{R}^{n \times n}$. The CS matrix in BCS is then 
\begin{equation}
         \mathbf{A} =        \begin{bmatrix}
        \textbf{A}_1 &\textbf{0}&\dots&\textbf{0}\\
        \textbf{0}&\textbf{A}_2&\ddots &\vdots\\
        \vdots & \ddots& \ddots & \textbf{0}\\
        \textbf{0}&\dots&\textbf{0}&\textbf{A}_\beta
        \end{bmatrix}
        \boldsymbol{\Pi},
    \label{eq:BCS_AEQ}
\end{equation}
where $\mathbf{A}_b \in \mathbb{R}^{\frac{m}{\beta} \times \frac{n}{\beta}}$. Setting $\boldsymbol{\Pi}$ to an identity matrix $\textbf{I}$ corresponds to the standard BCS scenario where the signal is partitioned into contiguous blocks. When the BCS matrix in \eqref{eq:BCS_AEQ} is used in \eqref{eq:CS_meas_model}, we observe that any measurement is a projection of a block within the sparse signal. We define $\textbf{z}=\boldsymbol{\Pi}\textbf{x}$. The $b^{\mathrm{th}}$ block of $\boldsymbol{\Pi} \textbf{x}$, defined as $\textbf{z}_b$, is a vector that contains entries of $\boldsymbol{\Pi} \textbf{x}$ indexed from $(b-1)n/\beta +1$ to $bn/\beta$. The dimension of the each signal block is $n/ \beta$, and $m/ \beta$ compressed measurements are acquired per block. 
\par In the BCS model, we define $\textbf{y}_b \in\R{\frac{m}{\beta}}$ as the measurements of the $b^{\mathrm{th}}$ block and $\textbf{v}_b$ as the associated noise. Then, $\textbf{y}=(\textbf{y}_1;\textbf{y}_2;\cdots;\textbf{y}_{\beta})$ is a column stacked version of the block measurements and $\textbf{v}=(\textbf{v}_1;\textbf{v}_2;\cdots;\textbf{v}_{\beta})$. The measurements in \eqref{eq:CS_meas_model} can be split in BCS as
\begin{equation}
    \textbf{y}_b =\textbf{A}_b  \textbf{z}_b+\textbf{v}_b \;\; \forall b \in \{1,2,\cdots, \beta\}.
\label{eq:BCS_eq}
\end{equation}
To estimate $\textbf{x}$, a straightforward approach is to first solve for $\{\textbf{z}_b\}_{b=1}^{\beta}$ from \eqref{eq:BCS_eq}. These problems can be solved in parallel. Then, the reconstructed blocks can be stacked together to obtain $\textbf{z}$. Finally, the permuntation operation in $\textbf{z}=\boldsymbol{\Pi}\textbf{x}$ can be inverted to estimate $\textbf{x}$.
\par Due to the block structure in BCS matrices, it is sufficient to store $\mathcal{O}(\beta {mn}/{\beta^2})$ entries of \eqref{eq:BCS_AEQ}. This requirement is $\beta$ times lower compared to the standard CS approach, which necessitates storing $\mathcal{O}({mn})$ entries. Furthermore, the complexity of matrix-vector multiplications in solving for one block in \eqref{eq:BCS_eq} is $\mathcal{O}({mn}/{\beta^2})$, when compared to $\mathcal{O}(mn)$ associated with \eqref{eq:CS_meas_model}. When the $\beta$ blocks are independently recovered, BCS requires a computational complexity that is $1/\beta^2$ lower than that of standard CS. When the $\beta$ blocks are estimated sequentially, BCS has a computational complexity that is $1/\beta$ lower than standard CS.
\subsection{Mutual coherence-based error bounds for BCS} \label{sec:bound_derivation}
In this section, we derive fundamental limits on parallel BCS-based recovery using the notion of mutual coherence. 
\par The mutual coherence of $\textbf{A}\in\R{m\times n}$ is defined as \cite{ben2010coherence}
\begin{equation}
        \mu(\textbf{A}) = \max_{ (k,\ell): k \neq \ell } \frac{|{\langle \textbf{a}_{k}, \textbf{a}_{\ell} \rangle}|}{{\|\textbf{a}_{k}\|_2 \|\textbf{a}_{\ell}\|_2}},
    \label{eq:coherence}
\end{equation}
where $\textbf{a}_{k}$ is the $k^{\mathrm{th}}$ column of $\mathbf{A}$. A small mutual coherence allows for better sparse reconstruction with the orthogonal matching pursuit (OMP)\cite{ben2010coherence}. For a CS matrix of size $m\times n$, the smallest possible mutual coherence that can be achieved is given by the Welch bound \cite{welch_bound}, i.e.,
\begin{equation}
    \mu(\textbf{A})\geq \sqrt{\frac{n-m}{m(n-1)}}.
    \label{eq:welch bound}
\end{equation}
The bound in \eqref{eq:welch bound} only depends on the size of the CS matrix.
\par We derive a new lower bound on the mutual coherence of the BCS matrix in \eqref{eq:BCS_AEQ}. To this end, we first notice that permuting the columns of $\textbf{A}$ does not impact the mutual coherence and assume $\boldsymbol{\Pi}=\textbf{I}$ in our analysis. Next, we observe that the columns of $\textbf{A}$ corresponding to different blocks are orthogonal. For instance, the first column of $\textbf{A}$ in \eqref{eq:BCS_AEQ} is orthogonal to all the columns indexed from $n/\beta +1$ to $n$, due to the block structure. Therefore, for a BCS matrix, the mutual coherence in \eqref{eq:coherence} is not determined by a pair of columns chosen from two different blocks. Applying the definition in \eqref{eq:coherence} for the BCS matrix in \eqref{eq:BCS_AEQ} results in 
\begin{equation}
\begin{aligned}
    \mu(\textbf{A}) &= \max \{ \mu(\textbf{A}_1), \mu(\textbf{A}_2), \cdots, \mu(\textbf{A}_{\beta})\}.
\end{aligned}
    \label{eq:welch_BCS}
\end{equation}
As each of the $\beta$ mutual coherences in \eqref{eq:welch_BCS} are lower bounded by the Welch bound associated with an $m/\beta \times n/ \beta$ matrix, we can write that 
\begin{align}
    \mu(\textbf{A}) \geq \sqrt{\frac{\frac{n}{\beta}-\frac{m}{\beta}}{\frac{m}{\beta}(\frac{n}{\beta}-1)}}=\sqrt{\frac{n-m}{m(\frac{n}{\beta}-1)}}.
      \label{eq:bound_BCS}  
\end{align}
For $\beta>1$, we observe that our lower bound in \eqref{eq:bound_BCS} for BCS is larger than the Welch bound in \eqref{eq:welch bound}. The key takeaway here is that BCS matrices cannot achieve the Welch bound in \eqref{eq:welch bound} due to the block diagonal constraint on the CS matrix.  For $\beta$ partitions, the smallest possible mutual coherence that BCS matrices can achieve is given by our bound in \eqref{eq:bound_BCS}. 
\par
\begin{figure}[h]
\centering
\includegraphics[trim=0cm 0cm 0cm 0.5cm, width=0.4\textwidth]{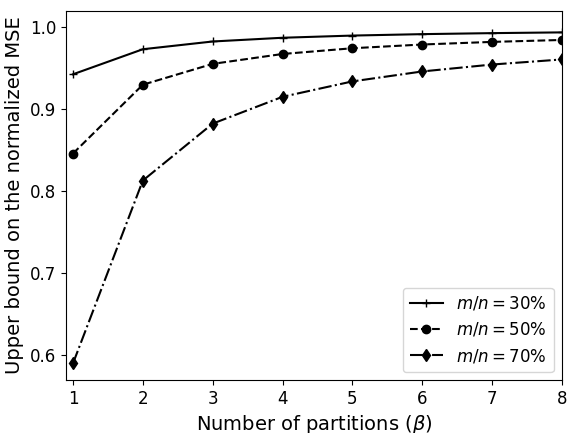}
\caption{\small The plot shows the upper bound on the MSE in \eqref{eq:omp_mu_bound} achieved by our coherence bound in \eqref{eq:BCS_AEQ}. The bound increases with $\beta$, the number of blocks. Here, $s=50$, $n=10^4$, $\alpha=0.5$ and $\sigma=10^{-2}$.}
\normalsize
  \label{fig:omp_e_bound}
\end{figure}

 The coherence bound in \eqref{eq:bound_BCS} can be used to find limits on the mean squared error (MSE) with the OMP algorithm. For an $s$-sparse signal, this error bound is given by \cite{ben2010coherence}
\begin{equation}
    \Vert\hat{\textbf{x}}-\textbf{x}\Vert_2^2\leq \frac{2(1+\alpha)}{(1-(s-1)\mu)^2}s\sigma^2\log{m},
    \label{eq:omp_mu_bound}
\end{equation}
where $\alpha$ is a constant independent of $m$ and $n$. We notice from \eqref{eq:omp_mu_bound} that a small mutual coherence results in a tight upper bound on the reconstruction error. When a BCS matrix in \eqref{eq:BCS_AEQ} is used for sparse recovery with the OMP, the tightest upper bound on the MSE is obtained when the mutual coherence is equal to the lower limit in \eqref{eq:bound_BCS}. A sketch of the MSE bound for OMP-based BCS is shown in Fig. \ref{fig:omp_e_bound} for different $\beta$. We observe that the error bound monotonically increases with $\beta$, indicating that partitioning the sparse recovery problem can lead to poor reconstruction.
\section{Data-driven serial BCS}
\begin{figure}[htbp]
\centering
\subfloat[Contiguous partitioning]{\includegraphics[trim=0cm 0cm 0cm 0cm,clip=true,width=3.25cm, height=2.7cm]{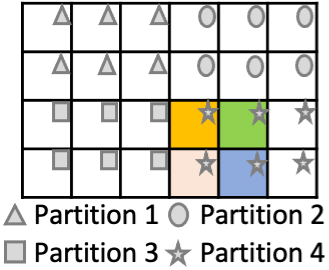}\label{fig:contin_part}}
\:\:
\subfloat[Comb-like partitioning]{\includegraphics[trim=0cm 0cm 0cm 0cm,clip=true,width=3.25cm, height=2.7cm]{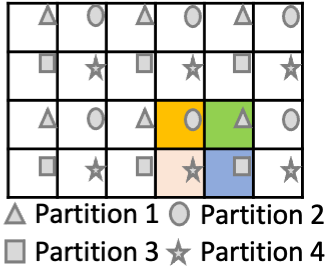}\label{fig:comb_part}}
\caption{\small Examples of partitioning strategies for $\beta=4$. The non-zero components of the 2D sparse signal here are shaded. For clustered sparse signals partitioned with a comb-like pattern, reconstructing one block provides side information on the support of other blocks.\normalsize}
\vspace{-5mm}
\label{fig:clustered_sparse}
\end{figure}

To alleviate the MSE loss due to partitioning in BCS, we exploit correlation across different blocks within the sparse signal during reconstruction. We demonstrate our approach by modeling and exploiting these correlations, using an example of a clustered sparse signal illustrated in Fig. \ref{fig:clustered_sparse}. 
\par A good partitioning strategy is key to the success of both standard BCS and our data-driven serial BCS. For instance, contiguous partitioning, shown in Fig. \ref{fig:contin_part}, that groups contiguous indices into a block does not work well for clustered sparse signals. This is because the non-zero coefficients are likely to be concentrated in a few blocks, resulting in poor reconstruction when an equal number of CS measurements are allocated per block. Comb-like periodic partitioning, shown in Fig. \ref{fig:comb_part}, is better suited to recover clustered sparse signals through BCS as it results in almost the same number of non-zero coefficients per block. In this paper, we consider clustered sparse signals because spatial wireless channels are known to exhibit such property in the Fourier dictionary \cite{group_sparse_mmwave}. Furthermore, comb-like partitions can be realized in hardware using the codes designed in \cite{comb_permutation}. We observe from Fig. \ref{fig:comb_part} that a non-zero entry reconstructed in a comb-like block (e.g., entry at $\Delta$) provides useful side-information on the support of its neighbours for clustered sparse signals. This information can be used as a prior to reconstruct other blocks.
\par Now, we discuss an offline algorithm to learn the side-information provided by a non-zero component on the support of its neighbours. Our algorithm considers clustered sparse signals and comb-like partitions. Using a dataset of sparse signals $\{\T{X}^{(j)}\}_{j=1}^J$, our algorithm estimates the probability that an immediate neighbour is non-zero conditioned on an entry being non-zero. This probability is determined empirically by counting the number of samples within the dataset in which the neighbouring entry is non-zero. For a 2D signal, a naive approach finds $\sim 8n$ such probabilities as there are $8$ neighbours around each entry except for those at the edges. For a $d^{\mathrm{th}}$ order tensor, this number increases to $\sim (3^d-1) n$. As storing all these probabilities consumes significant memory, we assume that the probability that a neighbour is non-zero is spatially invariant. Under this assumption, our algorithm only learns the support correlation kernel with $3^d-1$ entries for a $d^{\mathrm{th}}$ order tensor. Our procedure to learn this kernel $\T{\boldsymbol{\Theta}}$ from a dataset of sparse signals is summarized in Algorithm \ref{alg:kernel}. We explain this algorithm for the tensor case as we consider a fourth order tensor estimation problem in our simulations. 
\begin{algorithm}[h]
\caption{Construction of the support correlation kernel}
\label{alg:kernel}
\begin{algorithmic}[0]
\Statex \hspace{-4mm} \textbf{Input}: Sparse signals $\{\T{X}^{(j)}\}^J_{j=1}$ in $ \R{n_1\times n_2\times\dots\times n_d}$. 
\Statex \hspace{-4mm} \textbf{Define}: Set $\mathcal{S}_j$ has indices of non-zero entries in $\T{X}^{(j)}$.
\For{$j:1 \rightarrow J$}
\State Initialize $d^{\mathrm{th}}$ order kernel $\T{\boldsymbol{\kappa}}=\mathbf{0}\in\R{3\times 3 \times\cdots (d \, \mathrm{times})}$.
\For{index $(\omega_1,\omega_2, \cdots, \omega_d)$ in $\mathcal{S}_j$}
    \State Find immediate neighbours around this index:
    \State $\Omega=\{(\zeta_1,\zeta_2, \cdots, \zeta_d): |\omega_i-\zeta_i| \leq 1 \forall i \in [d]\}$.
    \State \textbf{Scan} through all the entries of $\T{X}^{(j)}$ at $\Omega$
    \State \hspace*{\algorithmicindent} \textbf{If} $\T{X}^{(j)}(\zeta_1,\zeta_2, \cdots, \zeta_d) \neq 0:$
    \State \hspace*{\algorithmicindent} Add $1$ to $\T{\boldsymbol{\kappa}}(\zeta_1-\omega_1,\zeta_2-\omega_2, \cdots, \zeta_d-\omega_d)$.
    \State Set $\T{\boldsymbol{\kappa}}(0,0,\cdots,0)=0$.
\EndFor
\State $\T{\boldsymbol{\Theta}}^{(j)}=\T{\boldsymbol{\kappa}}/{\text{cardinality}(\mathcal{S}_j)}$
\EndFor
\State\Return $\T{\boldsymbol{\Theta}} = \sum_{j=1}^J\T{\boldsymbol{\Theta}}^{(j)}/J$
\end{algorithmic}
\end{algorithm}
\begin{figure*}[h!]
\centering
\includegraphics[trim=0.6cm 0cm 0.6cm 0cm, width=0.9 \textwidth]{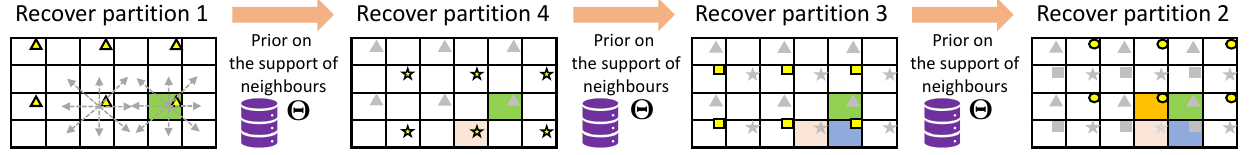}
\caption{\small Our proposed data-driven serial BCS method first reconstructs one of the signal blocks (partitions). Then, the reconstructed signal together with the learned support correlation $\T{\boldsymbol{\Theta}}$ is used to find a prior on the support associated with the other blocks. This prior is used together with the BCS measurements to reconstruct the next block, and the process is repeated until all the blocks are recovered.}
\normalsize
  \label{fig:summary_BCS_pipe}
\end{figure*}
\par We discuss notation involved in our data-driven BCS method to recover a sparse tensor $\T{X} \in \R{n_1\times n_2\times\dots\times n_d}$ from its compressed representation. Here, the dimension of the signal is $n=n_1n_2\cdots n_d$. Assuming that the $\ell^{\mathrm{th}}$ mode is partitioned by a factor of $\beta_{\ell}$, the total number of partitions is $\beta=\beta_1 \beta_2\cdots \beta_d$. The tensor $\T{X}_b \in \R{\frac{n_1}{\beta_1} \times \frac{n_2}{\beta_2}\times\dots\times \frac{n_d}{\beta_d}}$ comprises the entries of $\T{X}$ at the indices in partition $b$. Further, $\T{X}_{b,\mathrm{ext}}\in \R{n_1\times n_2\times\dots\times n_d}$ has the entries of $\T{X}$ at the indices in partition $b$ and zeros at all other indices. Finally, we define $\T{P}\in \R{n_1\times n_2\times\dots\times n_d}$ such that $\T{P}(\zeta_1,\zeta_2, \cdots, \zeta_d)$ is proportional to the probability that $\T{X}(\zeta_1,\zeta_2, \cdots, \zeta_d)$ is non-zero. After each stage of our algorithm that solves for a block within $\T{X}$, $\T{P}$ is updated using the reconstructed block and the support correlation kernel $\T{\boldsymbol{\Theta}}$.
\par Our data-driven BCS technique, summarized in Algorithm \ref{alg:proposed_BCS}, first solves for one of the $\beta$ blocks within the sparse signal from its compressed representation. Without loss of generality, our algorithm first reconstructs $\T{X}_1$, i.e., block $1$, from 
\begin{equation}
    \textbf{y}_1= \mathsf{A}_1 (\T{X}_1) + \mathbf{v}_1,
\end{equation}
where $\mathsf{A}_1(\cdot)$ denotes a linear compression operator. To solve for this first block, we set $\T{P}=s\mathbf{1}/n$, where $s$ is the average sparsity level estimated from the dataset. We use the logit-weighted OMP (LW-OMP) algorithm \cite{IT_paper_LWOMP}, that exploits the support prior $\T{P}$ together with the compressed measurements, to reconstruct a signal. The reconstructed block $\hat{\T{X}_1}$ is extended with zeros to obtain $\hat{\T{X}}_{1,\mathrm{ext}}$, which is then convolved with $\T{\boldsymbol{\Theta}}$ to update the support prior. The updated prior $\T{P}$ is then used to determine the block for which the most side-information on its support is available, i.e., the block for which $\mathrm{sum}\{\mathrm{vec}(\T{P}_b)\}$ is maximum. This block is recovered next and the procedure is repeated until all the blocks within $\T{X}$ are recovered. Our algorithm for the matrix case ($d=2$) is illustrated in Fig. \ref{fig:summary_BCS_pipe}.
\begin{algorithm}[h]
\caption{Proposed data-driven serial BCS algorithm}
\label{alg:proposed_BCS}
\begin{algorithmic}[0]
\Statex \hspace{-4mm} \textbf{Input}: BCS measurements $\{\textbf{y}_b\}^\beta_{b=1}$, CS operators $\{\mathsf{A}_b\}^\beta_{b=1}$,\\
\hspace{6.5mm} Learned support correlation kernel $\T{\boldsymbol{\Theta}}$.
\Statex \hspace{-4mm} \textbf{Define}: Set $\nu=[\beta]$, Support prior tensor  $\T{P}=s\mathbf{1}/n$ ($s$ is \\
\hspace{7mm} the average sparsity level), $\hat{\T{X}}=\mathbf{0}$ and $b=1$.
\For{ $j:1\rightarrow\beta$}
\vspace{1mm}
    \State $\hat{\T{X}}_{b} = \text{LW-OMP}(\textbf{y}_b,\mathsf{A}_b,\T{P}_b, \sigma^2)$ \hfill $\#\mathrm{Solve for block} b$
    \State $\hat{\T{X}}=\hat{\T{X}}+\hat{\T{X}}_{b, \mathrm{ext}}$
    \State $\T{P}=\T{P}+|\hat{\T{X}}_{b, \mathrm{ext}}\circledast\T{\boldsymbol{\Theta}}|$ $\#$Update prior using estimate
    \State $\nu \leftarrow \nu \setminus \{b\}$
    \State $b=\underset{k\in \nu}{\arg\max}[\,\mathrm{sum}\{\mathrm{vec}(\T{P}_k)\}]$  $\#$ Next block to solve
\EndFor
\State\Return $\Hat{\T{X}}$
\end{algorithmic}
\end{algorithm}
\par Our data-driven serial BCS with LW-OMP exploits structure across partitions but incurs higher computational complexity than standard BCS with OMP. This is because our method serially solves for the $\beta$ partitions, unlike standard BCS which can be parallelized. Further, support prior update in our method requires convolution, which adds a complexity of $\mathcal{O}(3^d n/ \beta)$ per partition. Considering $\beta$ partitions, the increase in complexity with our method over serial BCS is $\mathcal{O}(3^d n)$, still modest relative to unpartitioned standard CS.
\section{Simulation results}
\par We consider near-field spatial channel estimation between a $16\times 16$ transmitter (TX) and a $8 \times 8$ receiver (RX) at a carrier frequency of $300\, \mathrm{GHz}$. The TX and the RX use half-wavelength spaced uniform planar arrays. The RX is placed on a plane at a distance of $30\,\mathrm{cm}$ from the TX. We generate several channel realizations according to the propagation model in \cite{signal_paper}, by translating the RX along the plane and also rotating it at random. The kernel $\T{\boldsymbol{\Theta}}$ in our method was learned with Algorithm \ref{alg:kernel} using these realizations. Each channel realization is a complex-valued tensor in $\mathbb{C}^{16 \times 16 \times 8 \times 8}$. We use $\T{X}\in \mathbb{C}^{16 \times 16 \times 8 \times 8}$ to denote the 4D-discrete Fourier transform (4D-DFT) of the channel. As the channel exhibits clustered sparsity in the 4D-DFT, $\T{X}$ is a sparse tensor.
\begin{figure}[h!]
\centering
\includegraphics[trim=.5cm 0cm 0cm 1.3cm, width=0.48\textwidth]{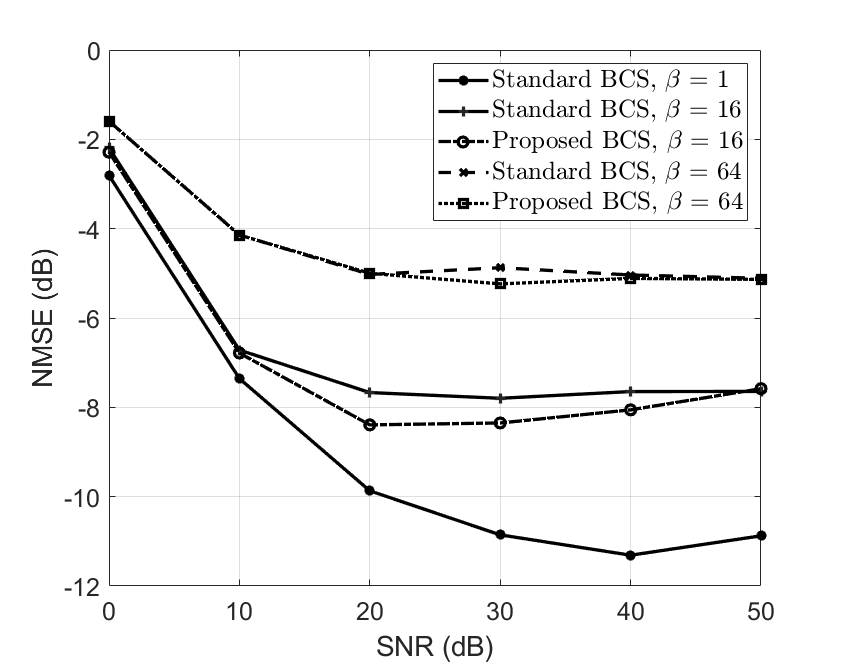}
\caption{\small NMSE with SNR for different partitions $\beta$. Here, the subsampling ratio is $40\%$. Our method outperforms standard BCS in the moderate SNR regime, at the expense of extra computations.}
\normalsize
  \label{fig:res_NMSE}
\end{figure}
\vspace{-2mm}
\par The measurements in BCS were acquired by applying 2D-codes in \cite{comb_permutation} at both the TX and the RX. These codes facilitate partitioning by implementing complementary comb-like masks over the sparse signal. By adjusting the periodicity within the comb, the number of partitions $\beta$ can be configured. In our simulations, we consider $\beta \in \{1, 16, 64\}$ to compare the performance of the proposed approach against standard BCS. Note that $\beta=1$ corresponds to the unpartitioned problem, which is solved using the standard OMP algorithm. Our approach employs the LW-OMP \cite{IT_paper_LWOMP} within Algorithm \ref{alg:proposed_BCS}, whereas standard BCS uses the classical OMP algorithm to solve for each partition. The transmit power is scaled to achieve the desired SNR in the compressed measurements. 
\begin{figure}[h!]
\centering
\includegraphics[trim=0.5cm 0cm 0cm 1.3cm, width=0.48\textwidth]{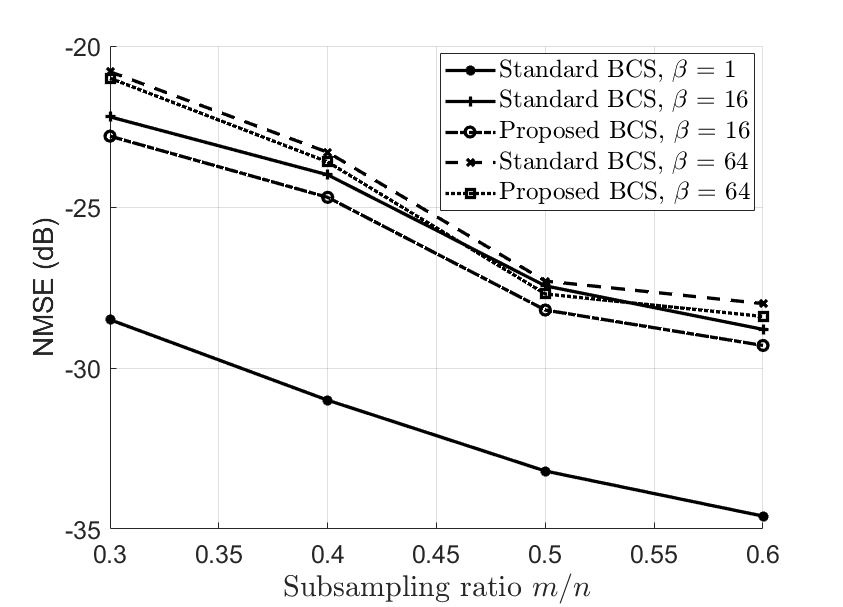}
\caption{\small For all the methods, NMSE monotonically decreases with the subsampling ratio for each $\beta$. Here, the SNR was set to $30\,\mathrm{dB}$. Our data-driven serial BCS method outperforms standard BCS.}
\label{fig:res_mu_f}
\end{figure}
\vspace{-2mm}
\par To evaluate the performance of BCS algorithms, we use the normalized mean squared error (NMSE) between the sparse channel and its estimate, i.e., $\mathbb{E}[\Vert \T{X}- \hat{\T{X}} \Vert_\mathrm{F}^2]/\mathbb{E}[\Vert \T{X} \Vert_\mathrm{F}^2]$, where the expectation is taken across all channel realizations. The dimension of the sparse vector solved for $\beta=1$ is $16\times 16 \times \times 8 \times 8 =16384$. With BCS, however, this dimension reduces to $16384/ \beta$ per partition. From Fig. \ref{fig:res_NMSE}, we observe that BCS yields higher NMSE than standard CS (equivalent to BCS with $\beta=1$). Moreover, the NMSE deteriorates for increasing $\beta$. These findings align with our analysis in Sec. \ref{sec:bound_derivation}, where higher values of $\beta$ led to an increased error bound.
\vspace{-1mm}
\begin{table}[h]
    \centering
    \begin{tabular}{|c|c|c|}\hline
    Number of blocks $\beta$  &  Standard BCS & Proposed method \\ \hline
     1  & 18.2 $\times$ $10^8$ ms& -\\ \hline 
     16 & 5910 ms& 9827 ms\\ \hline 
     64 & 476 ms & 1469 ms \\ \hline 
    \end{tabular}
    \caption{Computation time with standard BCS and the proposed method. BCS with $\beta=1$ is same as classical CS.}
    \label{tab:complexity}
    \vspace{-3mm}
\end{table}
\par We notice from Fig. \ref{fig:res_NMSE} that the proposed data-driven serial BCS outperforms standard BCS in the moderate SNR regime. At low SNR, the performance of both the methods is almost the same. This is because of the noise in the reconstruction, which impacts the support prior estimated, i.e., $\T{P}$, in Algorithm \ref{alg:proposed_BCS}. As the support prior estimate is not reliable at low SNR, it does not improve the reconstruction with LW-OMP. At high SNR, the measurements in each block provide reliable information to determine the support even without any prior information. Next, we observe from Fig. \ref{fig:res_mu_f} that the proposed method outperforms standard BCS for different subsampling ratios, i.e., $m/n$, at an SNR of $30\, \mathrm{dB}$. This performance improvement comes at the expense of an increased computational complexity compared to standard BCS, owing to the support prior calculation in our approach. The increase in complexity, however, is small when compared to the complexity of the unpartitioned CS problem for $\beta=1$. The computation times of all the methods, on a desktop computer, for $n=16384$ and a subsampling ratio of $\sim 40\%$ is summarized in Table \ref{tab:complexity}. 
\section{Conclusions}
In this paper, we studied block compressed sensing (BCS) for high-dimensional sparse recovery. We proved that the lower bound on the mutual coherence of a BCS matrix is higher than the Welch bound associated with a standard CS matrix of the same dimensions. We also proposed a data-driven serial BCS method, which uses reconstructed blocks to estimate priors on the support of subsequent blocks. Using simulations for the channel estimation problem, we showed that our method outperforms standard BCS at moderate SNR.
\bibliographystyle{IEEEtran}
\bibliography{References}
\end{document}